# On-surface synthesis of antiaromatic and open-shell indeno[2,1-*b*]fluorene polymers and their lateral fusion into porous ribbons

Marco Di Giovannantonio,[1,*] Kristjan Eimre,[1] Aliaksandr V. Yakutovich,[1] Qiang Chen,[2] Shantanu Mishra,[1] José I. Urgel,[1] Carlo A. Pignedoli,[1] Pascal Ruffieux,[1] Klaus Müllen,[2,3] Akimitsu Narita,[2,4,*] Roman Fasel[1,5,*]

[1]*Empa, Swiss Federal Laboratories for Materials Science and Technology, nanotech@surfaces Laboratory, 8600 Dübendorf, Switzerland*

[2]*Max Planck Institute for Polymer Research, 55128 Mainz, Germany*

[3]*Institute of Physical Chemistry, Johannes Gutenberg University Mainz, Duesbergweg 10-14, 55128 Mainz, Germany*

[4]*Organic and Carbon Nanomaterials Unit, Okinawa Institute of Science and Technology Graduate University, Okinawa 904-0495, Japan*

[5]*Department of Chemistry and Biochemistry, University of Bern, 3012 Bern, Switzerland*

**ABSTRACT**

Polycyclic hydrocarbons have received great attention due to their potential role in organic electronics and, for open-shell systems with unpaired electron densities, in spintronics and data storage. However, the intrinsic instability of polyradical hydrocarbons severely limits detailed investigations of their electronic structure. Here, we report the on-surface synthesis of conjugated polymers consisting of indeno[2,1-*b*]fluorene units, which are antiaromatic and open-shell biradicaloids. The observed reaction products, which also include a non-benzenoid porous ribbon arising from lateral fusion of unprotected indeno[2,1-*b*]fluorene chains, have been characterized *via* low temperature scanning tunneling microscopy/spectroscopy and non-contact atomic force microscopy, complemented by density-functional theory calculations. These polymers present a low band gap when adsorbed on Au(111). Moreover, their pronounced antiaromaticity and radical character, elucidated by *ab initio* calculations, make them promising candidates for applications in electronics and spintronics. Further, they provide a rich playground to explore magnetism in low-dimensional organic nanomaterials.

**INTRODUCTION**

Polycyclic aromatic hydrocarbons (PAHs) have received tremendous attention in recent decades due to their possible use as active layers in electronic devices. Depending on their size and geometry, they show diverse optical and chemical properties, but also offer various synthetic challenges.[1] Due to increased intrinsic reactivity, these challenges become serious in the case of antiaromatic and/or open-shell systems. This calls for special synthetic techniques and tools,[2,3] and in some cases alternative strategies with respect to traditional solution chemistry.[4–7] One such strategy is on-surface synthesis,[8] which enables the fabrication of (macro)molecular architectures that are difficult or impossible to realize in solution. On-surface synthesis has been extensively used for the fabrication of one- and two-dimensional (1D and 2D) materials, with a broad spectrum of chemical reactions to achieve covalently bonded structures under ultrahigh vacuum (UHV) conditions or at the solid-liquid or solid-vapor interface.[9–12] Careful design of the precursor molecules allows specific products to be achieved, permitting the fine tuning of the structural, electronic and magnetic properties. The most representative on-surface reactions are dehalogenative aryl-aryl coupling (Ullmann-type coupling)[13] and cyclodehydrogenation,[14] which have been applied to the fabrication of a variety of graphene-related materials, especially graphene nanoribbons (GNRs).[15] On-surface synthe-sis is also suited to fabricate intrinsically reactive molecular systems, due to the stabilization offered by the substrate and the UHV conditions. This advantage is of particular interest in dealing with antiaromatic or open-shell systems, and may contribute to establish new research opportunities in terms of chemical reactivity and unique electronic and magnetic properties in π-extended nanostructures.

Among the promising systems that would benefit from surface stabilization, indenofluorenes are of central importance and are emerging as carbon-based nanomaterials with intriguing properties.[16–23] They are non-alternant, non-benzenoid polycyclic hydrocarbons (PCHs) formed by a conjugated array of fused 6–5–6–5–6 rings with quinoidal bond pattern (see Scheme 1a). According to Hückel's rule, the presence of 20 π-electrons renders them anti-aromatic. Indenofluorenes may help to provide answers to fundamental questions about electronic configuration and reactivity in expanded, conjugated structures. Despite significant synthetic challenges, in the past few years derivatives of the five indenofluorene regioisomers have been synthesized in solution upon stabilization by protecting groups located at the most reactive sites, and characterized in detail.[16–23] Setting aside the recognition from organic chemistry, only a few works report indenofluorenes on surfaces,[24,25] although they offer unique opportunities for chemical and physical studies of highly reactive materials.

Given the potential of these molecules, it is also a challenge to target extended nano-arrays, *e.g.* polymers composed of indenofluorenes as repetitive units. Very recently, we described the on-surface synthesis and characterization of 1D polymers made from two of the five possible indenofluorene isomers, indeno[1,2-*b*]fluorene and indeno[2,1-*a*]fluorene.[24] However, the monomer design led to regioisomerism and failed to afford perfect structures. Further, the polymers were found to have a relatively large band gap of 2.3 eV and a closed-shell electronic configuration. Density-functional theory (DFT) calculations revealed that, among the five possible indenofluorene isomers, indeno[2,1-*b*]fluorene (IF) possesses one of the highest biradical character (y = 0.645).[26] A gain in aromatic sextets (from one to three) upon formal loss of a π-bond when transforming the closed- into the open-shell resonance structure provides an explanation to this phenomenon (Scheme 1a).[26,27] IF has a core structure of *meta*-quinodimethane (or *s*-indacene) which is highly reactive.[28] Tobe and co-workers achieved the synthesis and characterizations of IF using bulky mesityl groups at the five-membered ring apices for steric protection of radical sites.[19] IF was reported to have a singlet-triplet gap of -0.18 eV and a low electrochemical gap of 1.26 eV.[19] Theoretical studies have compared its local aromaticity with that of its isomers and also predicted promising optical properties.[26,29]

Here, we aim at the formation of 1D polymers made of pristine IF units, *i.e.* without any protecting group at the five-membered rings. Starting from 4,4"-dibromo-4',6'-dimethyl-1,1':3',1"-terphenyl as the monomer (**1** in Scheme 1b) on Au(111) under UHV conditions, we exploit the well-known dehalogenative aryl-aryl coupling to link the repeating units in a covalent fashion,[30–38] and use the methyl groups to selectively form five-membered rings *via* oxidative ring closure.[24] We demonstrate the successful on-surface synthesis of polymers consisting of unsubstituted IF as repeating units (*poly*-IF, **4/5** in Scheme 1b). These polymers are predicted to exhibit an open-shell singlet ground state by our spin-polarized DFT calculations of the system in gas phase. Unrestricted DFT and nucleus-independent chemical shift (NICS) calculations of *poly*-IF consisting of up to five repeating units show that the antiaromaticity and radical character of an isolated IF remain unperturbed upon polymerization. The reaction sequence – from dehalogenative aryl-aryl coupling to oxidative cyclization of methyl groups, and their step-wise dehydrogenation – is monitored by scanning tunneling microscopy/spectroscopy (STM/STS) and non-contact atomic force microscopy (nc-AFM), supported by DFT calculations. Our results not only demonstrate the synthesis of antiaromatic and open-shell polymers potentially useful for electronics and spintronics, but also provide a basis for the fundamental understanding of polyradical systems.[39]

**Scheme 1. Indenofluorene isomers and on-surface reaction scheme towards *poly*-IF.** (a) Resonance structures of indenofluorene isomers in closed- and open-shell forms, with Clar's sextets highlighted in blue. Ground state electronic configurations and biradical character indices (y) are reported. The latter have been calculated at the LC-UBYLP/6-311+G** level of theory. Values for structures *I-IV* are taken from Ref. [26]. In this work, we confirm the y obtained for structure *IV* and calculate it for *V*. (b) Synthetic strategy to obtain polymers consisting of IF units (*poly*-IF). The on-surface reaction temperatures are $T_1$=100-150 °C, $T_2$=200-250 °C and $T_3$=350-400 °C. Structures 4 and 5 represent the closed- and open-shell resonance forms of *poly*-IF, respectively. Porous ribbons 6 are initially formed at a temperature of $T_4$=300-350 °C, where they coexist with 3.

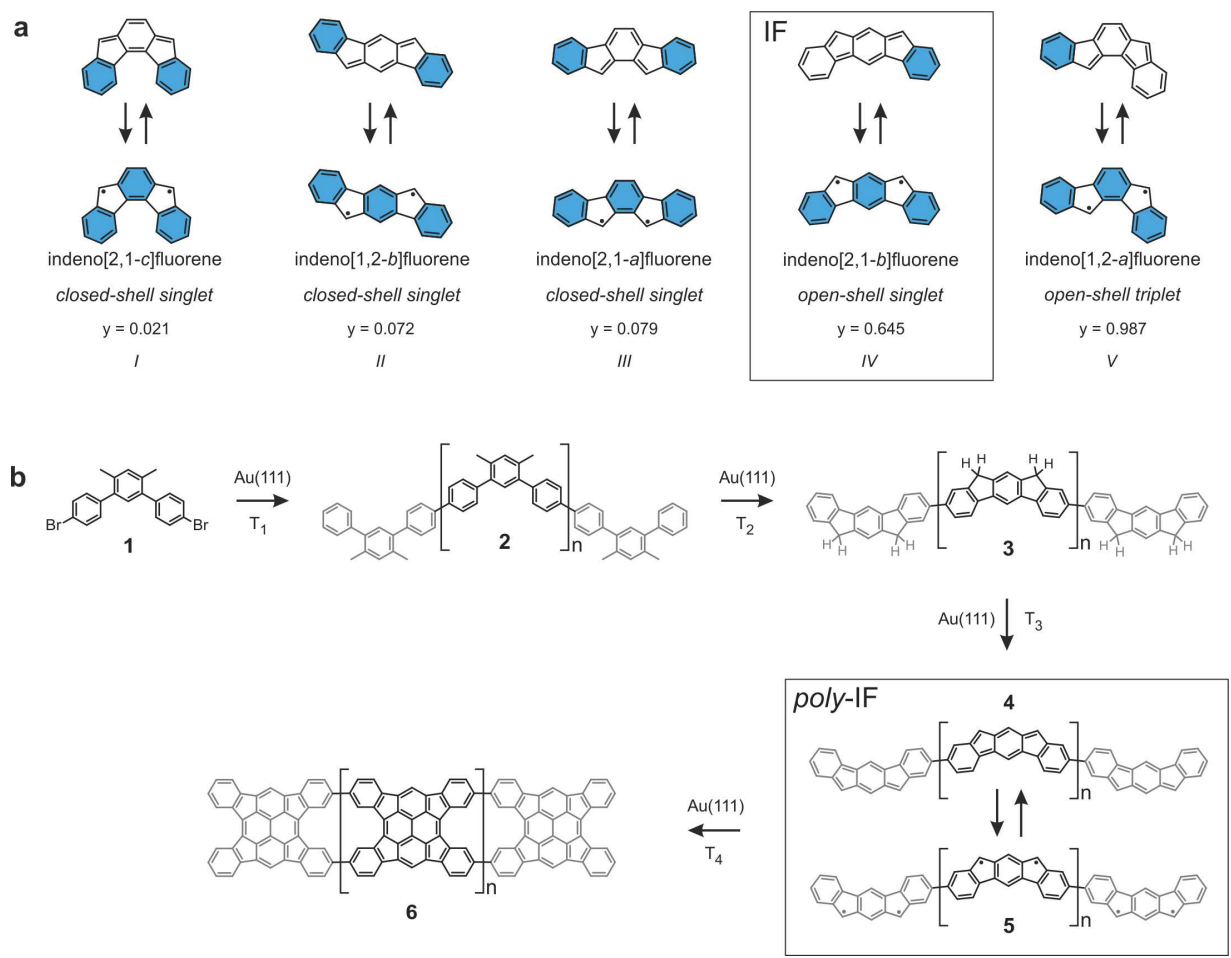

**METHODS**

**STM/STS and nc-AFM experiments.**

The on-surface synthesis experiments were performed under ultrahigh vacuum (UHV) conditions with base pressure below $2\times10^{-10}$ mbar. Au(111) substrates (MaTeck GmbH) were cleaned by repeated cycles of $Ar^+$ sputtering (1 keV) and annealing (470 ºC). The precursor molecules were thermally evaporated onto the clean Au(111) surface from quartz crucibles heated at 60 ºC with a deposition rate of ~ 0.5 Å·min$^{-1}$. STM images were acquired with a low-temperature scanning tunneling microscope (Scienta Omicron) operated at 5 K in constant-current mode using an etched tungsten tip. Bias voltages are given with respect to the sample. Constant-height dI/dV spectra and maps were obtained with a lock-in amplifier (f = 610 Hz). Nc-AFM measurements were performed at 5 K with a tungsten tip placed on a QPlus tuning fork sensor.[40] The tip was functionalized with a single CO molecule at the tip apex picked up from the previously CO-dosed surface.[41] The sensor was driven at its resonance frequency (27043 Hz) with a constant amplitude of 70 pm. The frequency shift from resonance of the tuning fork was recorded in constant-height mode using Omicron Matrix electronics and HF2Li PLL by Zurich Instruments. The Δz is positive (negative) when the tip-surface distance is increased (decreased) with respect to the STM setpoint at which the feedback loop is open.

**Computational details.**

The equilibrium geometries of the molecules adsorbed on the Au(111) surface and the corresponding STM images were calculated with the CP2K code[42] implementing DFT. We used the PBE functional with addition of van der Waals corrections. The surface/adsorbate systems were modeled with a slab consisting of 4 atomic layers of Au. Geometry optimizations were performed by keeping the bottom two layers of the slab fixed to ideal bulk positions. For the simulation of the AFM images, we used an empirical model mimicking the CO functionalized tip *via* two probe particles. The natural orbital occupations and NICS$_{zz}$ were calculated at the LC-UBLYP/6-311+G** level of theory for the isolated oligomers optimized at the U(R)B3LYP/6-311G** level of theory. For complete details, see Supporting Information Section 2.

**RESULTS AND DISCUSSION**

The on-surface reactions affording indenofluorene-based polymers are studied by depositing a submonolayer coverage of monomer **1** onto a clean Au(111) surface held at room temperature (RT), under UHV conditions. High-resolution STM (Figure 1a, top) and constant-height frequency shift nc-AFM imaging of an isolated molecule (performed with a CO-functionalized tip; Figure 1a, bottom) reveal the presence of bromine atoms still attached to the central backbone, confirming that **1** adsorbs intactly on the surface. The monomers self-assemble into close-packed islands due to Br⋯H interactions, as shown in Figure 1b.

Annealing the substrate to 150 ºC activates dehalogenative aryl-aryl coupling. Upon debromination, the monomers covalently link to each other, resulting in the formation of zigzag polymeric chains (Figure 1c). The detached bromine atoms, visible in between the polymers, are chemisorbed on the gold substrate. As a result, they alter the Au(111) (22×√3) surface reconstruction, and mediate the chain packing *via* Br⋯H interactions (Figure 1d). The appearance of alternating bright protrusions in the chains indicates that the methyl groups are still unreacted, reflecting the chemical structure of the expected intermediate product **2**.

Further annealing of the surface to 250 ºC produces a remarkable change in the appearance of the polymers (see Supporting Figure S1, where large scale STM images at all the investigated temperatures are reported). The Au(111) (22×√3) surface reconstruction is now restored, and the polymers, no longer assembled into islands, meander over the substrate (Figure 1e). These observations suggest that bromine atoms have desorbed from the gold substrate, as is ex-pected at this annealing temperature.[43] The nc-AFM image in Figure 1g reveals the structural details of a polymer segment. Each unit is composed of a sequence of fused 6–5–6–5–6 rings, indicating that the cyclization of the methyl groups toward the neighboring aromatic rings has occurred.[24] Noticeably, the apices of all five-membered rings appear brighter. This suggests that while one hydrogen atom has been removed from each methyl group in the cyclization process, two of them are still connected to the carbon atom, in agreement with previous studies reporting the brighter imaging of 2H-functionalized carbon atoms by nc-AFM.[25,44] Based on this assumption, we simulated the nc-AFM image (Figure 1h) of the suggested structure (Figure 1i), which agrees very well with the experimental image. This finding confirms that polymers **3**, consisting of 10,12-dihydroindeno[2,1-*b*]fluorene (2H-IF) units, are formed at this stage (*poly*-2H-IF, **3**). Height-dependent experimental and simulated nc-AFM images of the structure in Figure 1i are reported in Supporting Figure S2, demonstrating that the bright protrusions due to the 2H moieties are more evident at larger tip-molecule distances.

A voltage-dependent differential conductance spectrum (dI/dV *vs.* V) acquired on **3** reveals peaks in the density of states (DOS) at −1.1 V and +2.6 V (Figure 1j), corresponding to tunneling through the conduction and valence bands (CB and VB), respectively. Spatial mapping of the dI/dV signal (dI/dV maps) at the peak positions reveals excellent correspondence with the DFT-computed LDOS maps (Supporting Figure S3) and corroborates the assignment of the conductance peaks to orbital resonances. Thus, **3** exhibits a large band gap of 3.7 eV on Au(111).

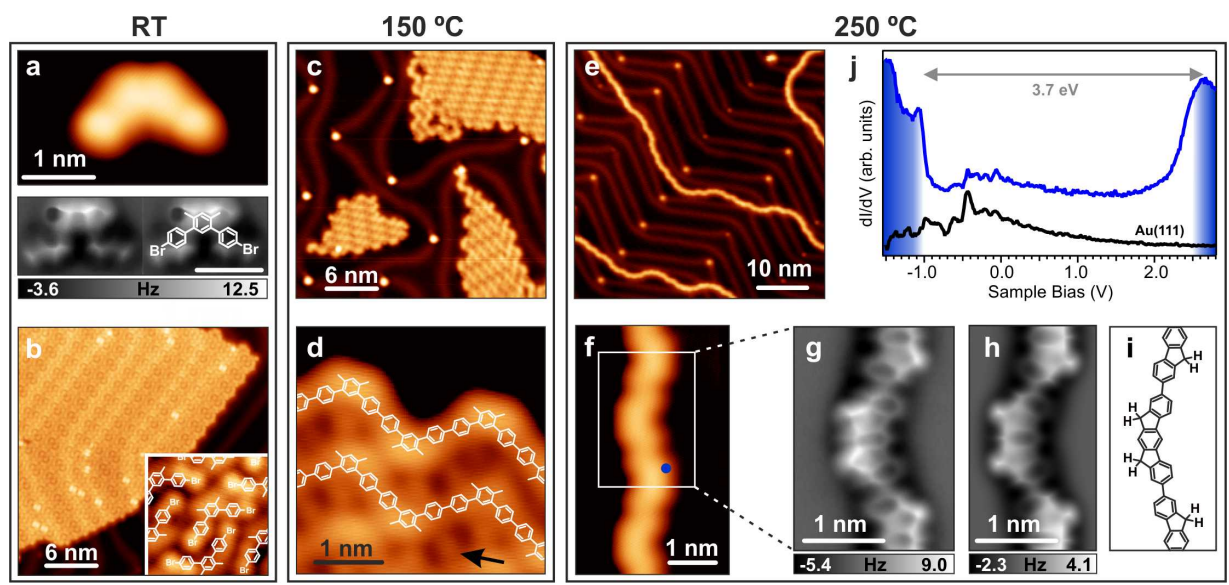

**Figure 1.** From **1** to *poly*-2H-IF (**3**) on Au(111). (a) High-resolution STM (top, $V_b$ = –0.30 V, $I_t$ = 50 pA) and nc-AFM (bottom, scale bar is 1 nm, $\Delta z$ is +2.5 Å with respect to STM set point: –0.005 V, 100 pA) images of **1** after deposition on Au(111) held at RT. (b) Large-scale STM image of a self-assembled island of **1** ($V_b$ = –1.50 V, $I_t$ = 50 pA). The inset shows a magnified area with superimposed molecular models ($V_b$ = –1.50 V, $I_t$ = 150 pA). (c) STM image of the surface after annealing at 150 ºC for 10 min ($V_b$ = –1.00 V, $I_t$ = 10 pA). (d) Magnified area from panel c with superimposed polymer models ($V_b$ = –0.90 V, $I_t$ = 20 pA). The black arrow indicates bromine atoms. (e) STM image of the surface after annealing at 250 ºC for 10 min ($V_b$ = –1.50 V, $I_t$ = 10 pA). (f) Magnified area from panel e ($V_b$ = –0.05 V, $I_t$ = 100 pA). (g) nc-AFM image acquired at the indicated segment of the polymer ($\Delta z$ = +2.1 Å with respect to STM set point: –0.005 V, 100 pA). (h) Simulated nc-AFM image for the chemical structure of *poly*-2H-IF in panel i. (j) dI/dV spectrum acquired on the polymer at the position indicated by the blue dot in panel f (blue curve), and reference spectrum taken on the bare Au(111) surface (black curve). The spectra are vertically shifted for clarity.

**Lateral fusion into porous ribbons**

To further dehydrogenate the newly formed five-membered rings of **3** toward the targeted $sp^2$-conjugated IF polymer **4**, the substrate was annealed at 310 ºC. Figure 2a displays a representative STM image of the resulting phase. While no morphological changes are observed in single chains (which are still of type **3**), we observe the lateral fusion of some polymers upon dehydrogenative C-C coupling. This interchain fusion generates ribbons hosting uniform pores with a rim composed of 16 carbon atoms (referred to as **6**, see nc-AFM image in Figure 2b). The average length of **6** is 10 nm, with a maximum observed length of 30 nm. Around 21% of the molecular units (out of more than 3000) form the ribbons **6**, while the rest remains in chains of type **3**.

Dehydrogenation of the polymeric chains towards the formation of **4** (activation barrier of 1.62 eV, as from Supporting Figure S6a) could be sufficient to initiate the lateral fusion into ribbons, given the low energy barrier needed for such a coupling (0.56 eV, Supporting Figure S6c). However, because the polymers present at this stage of the reaction are still of type **3**, the formation of **6** has to proceed *via* an alternative pathway with a lower activation barrier. To investigate this possibility, we performed calculations for the dehydrogenation of molecular units on Au(111) in proximity to a gold adatom (Supporting Figure S6b). In this scenario, the dehydrogenation of 2H-IF into IF requires only 1.52 eV. We know that this mechanism can only take place if the residence time of the gold adatoms next to the chains is long enough, given the high mobility of such adatoms on the (111) surface (diffusion barrier of 0.22 eV).[45] We speculate that this criterion is fulfilled when an adatom is confined between two polymers of type **3** close to each other, providing a route to their lateral fusion *via* dehydrogenation. To exclude other possible mechanisms, we also performed DFT calculations of merging processes of an IF and a 2H-IF unit as well as two 2H-IF units, using constrained geometry optimization approaches (Supporting Figure S6d and S6e). These calculations resulted in unsought products with higher energy barriers, supporting the Au adatom-mediated mechanism.

dI/dV spectra acquired on **6** (Figure 2c) reveal peaks in the DOS at −1.5 V and +0.7 V. The assignment of the peaks to frontier orbital resonances is confirmed by the qualitative agreement between the experimental constant height dI/dV maps acquired at the energetic positions of the peaks and the DFT-computed LDOS maps of the CB and VB of **6** (Supporting Figure S3d-g). The band gap of **6** on Au(111) thus amounts to 2.2 eV.

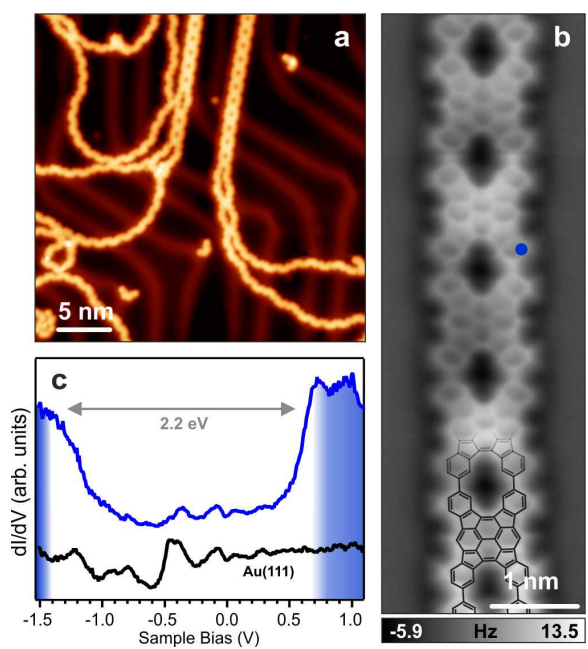

**Figure 2.** Porous ribbons (**6**) formed *via* lateral fusion on Au(111). (a) STM image of the surface after annealing at 310 ºC ($V_b$ = –0.7 V, $I_t$ = 20 pA). Newly formed segments of **6** coexist with polymeric chains **3**. (b) Constant-height frequency-shift nc-AFM image of a porous ribbon segment ($\Delta z$ = +2.4 Å with respect to STM set point: –0.005 V, 150 pA). (c) dI/dV spectrum acquired on **6** at the position indicated by the blue dot in panel b (blue curve), and reference spectrum taken on the bare Au(111) surface (black curve).

**Formation of *poly*-IF**

Annealing of the surface to 360 ºC produces partial changes in the morphology of single chains, which are almost complete at 410 ºC (Figure 3a). The nc-AFM image in Figure 3b and the corresponding scheme in Figure 3c reveal a clear chemical transformation of **3**. Some flat units composed of fused 6–5–6–5–6 rings are visible, and do not present any bright protrusion at the apices of the five-membered rings. This modification indicates the removal of a single hydrogen atom from the five-membered ring apices of **3** (*i.e.* from $CH_2$ to CH), thus leading to the formation of *poly*-IF. Apart from flat units in *poly*-IF we also observe some tilted ones, which occur when the carbon atoms at the apices of five-membered rings lie on top of a gold atom of the Au(111) surface. Figures 3d,e show the central unit of a DFT-optimized structure of an IF trimer on Au(111). The carbon atoms indicated by the red arrows lie nearly on top of Au atoms of the first substrate layer. The interaction between the molecule and the surface results in a slight upward displacement of the involved Au atoms (green arrows in Figure 3e) and a deviation from planarity of the hydrogen atoms linked to these carbon atoms (blue arrows in Figure 3e). If the carbon atoms at the apices of the five-membered rings are not on top of Au atoms, the IF unit remains nearly flat, as demonstrated by the DFT-optimized geometry in Figure 3f,g. Experimental high-resolution nc-AFM images of the two types of units (Figure 3h,j) are well-reproduced by the simulated images of the two geometries discussed above (Figure 3i,k) which confirms our interpretation. We note that due to the intrinsic flexibility of these polymers, it is easier for a unit to find an adsorption configuration with tilted geometry, while flat units are correspondingly rarer. More details on the origin of the two different configurations are reported in Supporting Figure S5. There, we present charge density difference plots for both cases, which reveal that in the case of tilted IF units a chemisorption contribution to the polymer-substrate interaction emerges. Moreover, we exclude that the tilted units could host carbene functionalities (formed *via* complete dehydrogenation of the carbon atoms at the five-membered rings' apices), as the energy barrier to complete dehydrogenation is too high to be accessible at the annealing temperatures applied in experiment (Supporting Figure S6a).

The IF polymers coexist with an increased amount of **6** due to enhanced lateral fusion at higher temperatures: 35% (43%) of the monomers at 360 ºC (410 ºC) form **6**. It must be noted that the dehydrogenation from 2H-IF to IF occurs in a wide temperature window. While the ratio between 2H-IF and IF in the polymeric chains is about 3:1 at 360 ºC, IF is the majority species at 410 ºC (see Supporting Figure S7). Also, not every chain is laterally fused into **6** because of the limited chain mobility on the surface after coupling reactions pinning their ends.

To investigate the electronic properties of the indeno[2,1-*b*]fluorene polymers we performed STS measurements on a flat unit (we could not obtain clear spectroscopic signals on a tilted unit, presumably due to increased interaction with the substrate). We detected peaks in the dI/dV spectrum at –0.2 V and +0.2 V (Figure 3l). Constant-height dI/dV maps at these energies match the calculated LDOS maps (Figure 3m-p). This demonstrates the successful formation of a polymer with exceedingly low band gap of 0.4 eV on Au(111). The trend of experimental band gaps measured on Au(111) for **3**, **4** and **6** is reproduced well by the corresponding DFT-calculated gap values, reported in Supporting Figure S4.

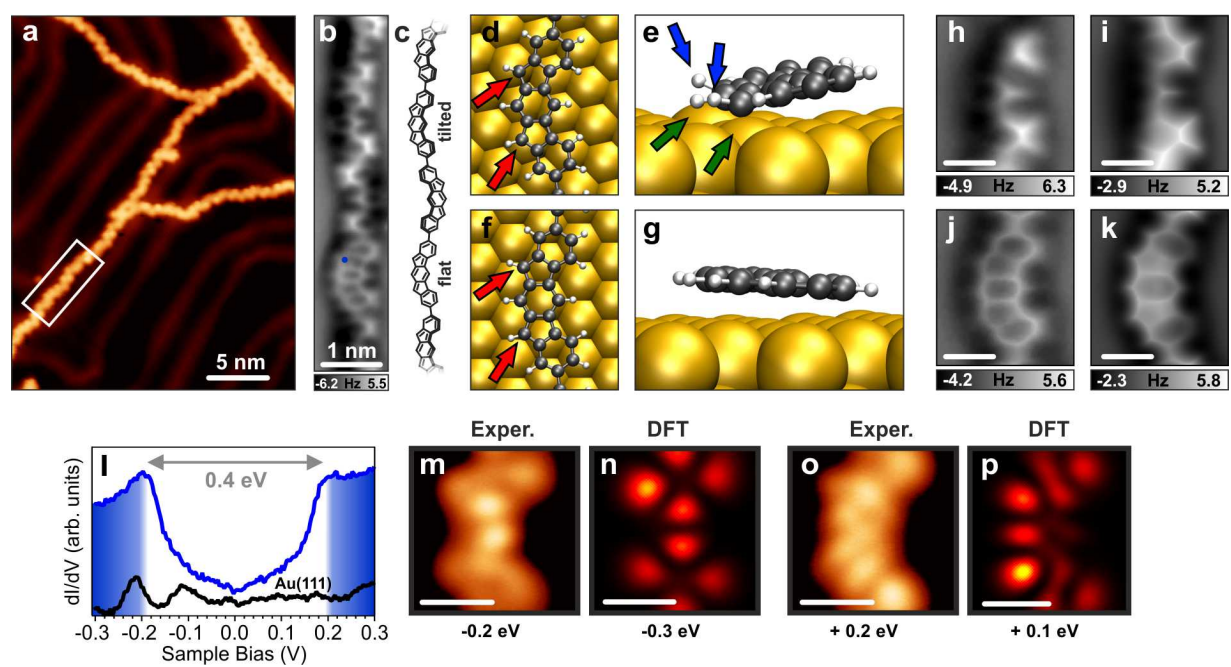

**Figure 3.** *Poly*-IF on Au(111). (a) STM image of the surface after annealing to 410 ºC for 10 min ($V_b$ = –0.3 V, $I_t$ = 30 pA). (b) nc-AFM image of the segment highlighted by the white rectangle in panel a (Δz = +1.6 Å with respect to STM set point: –0.005 V, 50 pA). (c) Molecular scheme of the structure reported in panel b. (d-g) Top and side views of the DFT optimized geometries of oligomers where the carbon atoms at five-membered rings' apices (red arrows) lie on top of Au atoms (d,e) or on hollow sites (f,g). In the first case, the interaction of the molecule with the gold lifts up two Au atoms (green arrows) and induces a $sp^3$ configuration, resulting in the corresponding hydrogen atoms pointing up (blue arrows). In the second case, the resulting geometry is flat and all the hydrogen atoms are in plane. (h-k) Experimental (h,j) and simulated (i,k) nc-AFM images of a tilted (h,i) and flat (j,k) polymer unit. Scale bars: 0.5 nm. Δz = +1.7 Å (h) and +1.5 Å (j) with respect to STM set point: –0.005 V, 50 pA. (l) dI/dV spectrum acquired on *poly*-IF at the position indicated by the blue dot in panel b (blue

curve), and reference spectrum taken on the bare Au(111) surface (black curve). (m-p) Constant-height experimental and theoretical dI/dV maps of a flat IF unit at the indicated voltages. Scale bars: 0.5 nm.

**Theoretical characterization of the electronic structure of *poly*-IF**

Accessing the magnetic structure of open-shell systems with scanning probe techniques represents a challenge due to the extremely weak magnetic anisotropy of graphene-related structures.[46] Therefore, we performed DFT calculations at the (U)B3LYP/6-311G** level of theory for an IF pentamer in gas phase to elucidate its electronic ground state. Our calculations reveal that the open-shell singlet state is more stable than the closed-shell state by 0.42 eV. Moreover, the spin density of the open-shell singlet state is mainly located at the apices of five-membered rings (Figure 4a), in line with Clar's theory. The geometry of the gas phase pentamer, optimized at the UB3LYP/6-311G** level, indicates that every other unit is twisted with respect to its neighbors such that the dihedral angle of the unit is ~35°.

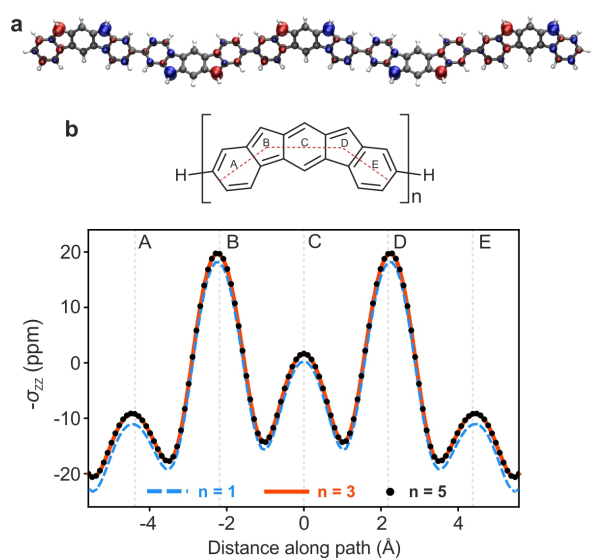

**Figure 4.** Theoretical electronic characterization of *poly*-IF. (a) Spin density isovalues at ±0.01 e·(a.u.)$^{-3}$ calculated at the LC-UBLYP level of theory. (b) Out-of-plane nucleus-independent chemical shift NICS$_{zz}$-1 (lower panel) calculated at the GIAO-LC-UBLYP/6-311+G** level of theory along the path indicated in the upper panel for an isolated monomer (n=1) and the central unit of a trimer (n=3) and a pentamer (n=5), respectively.

**Table 1. Computed radical character of *poly*-IF.**

|  | Monomer | Trimer | Pentamer |
|---|---|---|---|
| Natural orbital occupations | $n_{LUNO} = 0.645$ | $n_{LUNO} = 0.655$ | $n_{LUNO} = 0.657$ |
|  | $n_{LUNO+1} = 0.009$ | $n_{LUNO+1} = 0.645$ | $n_{LUNO+1} = 0.652$ |
|  | $n_{LUNO+2} = 0.009$ | $n_{LUNO+2} = 0.636$ | $n_{LUNO+2} = 0.645$ |
|  | $n_{LUNO+3} = 0.008$ | $n_{LUNO+3} = 0.009$ | $n_{LUNO+3} = 0.638$ |
|  | $n_{LUNO+4} = 0.006$ | $n_{LUNO+4} = 0.009$ | $n_{LUNO+4} = 0.634$ |
| $\dfrac{\sum_i (1 - |1 - n_i|)}{N_{units}}$ | 1.368 | 1.362 | 1.361 |

The natural orbital occupation numbers were obtained within the unrestricted density functional theory[47] at the LC-UBLYP/6-311+G** level of theory. The number of odd electrons per monomer (bottom row) was estimated from the occupation numbers with the expression proposed by Head-Gordon.[48]

To characterize the degree of antiaromaticity of *poly*-IF, we performed nucleus-independent chemical shift (NICS)[49] calculations on the monomer and on the central unit of a trimer and a pentamer (Figure 4b). The GIAO-LC-UBLYP/6-311+G** NICS$_{zz}$ method was used to find the magnetic shielding tensor component (-σ$_{zz}$) values, calculated at 1.0 Å above the molecular plane defined by the central units along the indicated trajectory. Negative values of -σ$_{zz}$ correspond to aromaticity of the system, while positive values indicate antiaromaticity. Figure 4b reveals that the five-membered rings are clearly antiaromatic and the six-membered rings at the edges are aromatic, while the central six-membered ring possesses values close to zero, typical of non-aromaticity. Compared to the monomer, the trimer and pentamer show a slight increase in antiaromaticity. Both the trimer and pentamer give matching values, indicating a converged result representative of the NICS values of a polymer. The overall similarity of the calculated NICS scans suggests that the repetition of the IF scaffold *via* C-C bonds does not alter the aromatic/antiaromatic properties. Therefore, we can safely assume *poly*-IF to maintain the antiaromaticity of an isolated IF.

The radical character of an organic compound is an important value to describe its open- or closed-shell electronic configuration. To provide a description of the radical character of *poly*-IF, we performed an analysis of natural orbital occupation numbers based on unrestricted DFT[47] calculations at the LC-UBLYP/6-311+G** level of theory for the monomer, trimer and pentamer. The results are summarized in Table 1. The calculated biradical character of the monomer (occupation number of LUNO) is 0.645, in agreement with previous work.[26] To compare the radical character of the biradical monomer, hexaradical trimer and decaradical pentamer, we estimated the total number of unpaired electrons in each system with the expression proposed by Head-Gordon[48] and divided it by the number of units in the oligomer. The obtained values are 1.368, 1.362 and 1.361, respectively, showing that the radical character is unaffected upon growth into longer oligomers.

**CONCLUSIONS**

We have demonstrated the Au(111)-assisted synthesis of polymers made of unprotected indeno[2,1-*b*]fluorene units, so far only studied as isolated monomers in solution after stabilization by bulky protecting groups. We monitored all the consecutive reaction steps and identified the temperature intervals of methyl group cyclization (around 200 ºC) and further dehydrogenation from $CH_2$ to CH (around 370 ºC). All observed structures could be unambiguously characterized by means of nc-AFM imaging. We also unraveled the formation of an unexpected reaction product formed by cross-dehydrogenative coupling (lateral fusion) of two IF polymers – a porous ribbon consisting of repeating tetraindenopyrene motifs, with a band gap of 2.2 eV. The IF polymer adsorbed on Au(111) is characterized by a low band gap of 0.4 eV. DFT calculations of a free-standing model oligomer reveal a significant open-shell character with spin densities predominantly located at the apices of five-membered rings, in agreement with Clar's theory. Additionally, we demonstrated that the obtained *poly*-IF retains the antiaromaticity and radical character of an isolated IF unit. The presence of multiple spins in IF polymers makes them promising components for carbon-based spintronic circuits. Finally, they are ideal systems to investigate the interplay of antiaromaticity and open-shell character, a phenomenon rarely studied in organic chemistry.[50]

## ASSOCIATED CONTENT

**Supporting Information**

The Supporting Information is available free of charge on the ACS Publications website at DOI: …

Precursor synthesis and characterizations, additional computational details, additional experimental and theoretical results (PDF)

## AUTHOR INFORMATION

**Corresponding Authors**

*E-mail: marco.digiovannantonio@empa.ch

*E-mail: narita@mpip-mainz.mpg.de

*E-mail: roman.fasel@empa.ch

**ORCID**

Marco Di Giovannantonio: 0000-0001-8658-9183

**Notes**

The authors declare no competing financial interest

## ACKNOWLEDGMENTS

This work was supported by the Swiss National Science Foundation, the NCCR MARVEL funded by the Swiss National Science Foundation (51NF40-182892), the European Union's Horizon 2020 research and innovation program under grant agreement number 785219 (Graphene Flagship Core 2), and the Office of Naval Research

BRC Program. Computational support from the Swiss Supercomputing Center (CSCS) is gratefully acknowledged. We are thankful to Lukas Rotach (Empa) for his excellent technical support during the experiment, and to Dieter Schollmeyer (Institute for Organic Chemistry, Johannes Gutenberg University Mainz) for single-crystal X-ray structural analysis.

Graphical abstract:

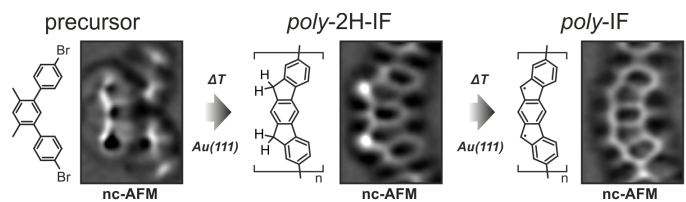

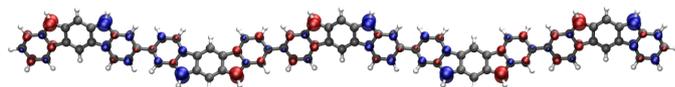

*poly*-IF: antiaromatic, open-shell